\documentclass[amsmath,amssymb,amsfonts,aps,prl,superscriptaddress,bibnotes,showpacs,showkeys,longbibliography,twocolumn]{revtex4-2}

\usepackage[english]{babel}
\usepackage{graphicx}
\usepackage{hyperref}
\usepackage{xcolor}
\usepackage{bbm}
\usepackage{amsthm}

\DeclareMathOperator{\sech}{sech}
\newcommand{\me}{\mathrm{e}}

\AtBeginDocument{%
    \newwrite\bibnotes
    \def\bibnotesext{Notes.bib}
    \immediate\openout\bibnotes=\jobname\bibnotesext
    \immediate\write\bibnotes{@CONTROL{REVTEX41Control}}
    \immediate\write\bibnotes{@CONTROL{%
    apsrev41Control,author="08",editor="1",pages="1",title="0",year="1"}}
     \if@filesw
     \immediate\write\@auxout{\string\citation{apsrev41Control}}%
    \fi
}%

\begin{document}
\author{Jonas Berx}
\email[]{jonas.berx@nbi.ku.dk}
\affiliation{Niels Bohr International Academy, Niels Bohr Institute, University of Copenhagen, Blegdamsvej 17, 2100 Copenhagen, Denmark}
\title{Geometry and universal scaling of Pareto-optimal signal compression}

\begin{abstract}
    I investigate the generic problem of lossy compression of a fluctuating stochastic signal $X$ into a discrete representation $Z$ through optimal thresholding. The signal modulates transition rates of a two-state system described by a binary variable $Y$. Optimising the retained mutual information between $Z$ and $Y$ under a constraint on fixed encoding cost of $Z$ reveals Pareto-optimal trade-offs, determined numerically using genetic algorithms. In the small-noise regime, these fronts are either concave or exhibit piecewise convex ``intrusions'' separated by first-order transitions in the optimal protocol. An analytical high-rate expansion shows that the optimal threshold density follows a universal cube-root scaling with the product of the prior distribution and the Fisher information associated with the response, which holds qualitatively even for few discrete states. Extending the analysis to non-Gaussian fluctuations reveals that for some parameters optimal encoders can yield strictly better information–cost trade-offs than Gaussian surrogates, meaning the same information content can often be achieved with fewer discrete readout states.
\end{abstract}
\date{\today}

\maketitle


What is the best way of reducing the compression cost of data $X$ while retaining as much information about another quantity $Y$ in the process? A system that is confronted with such a problem must generally solve some version of the Information Bottleneck (IB)~\cite{tishby2000informationbottleneckmethod}, i.e., trade-offs between accuracy and cost. Such trade-offs are central in a myriad of biochemical systems involved in sensing or signalling~\cite{Tottori2025,Tottori2025_long}, as well as in machine learning~\cite{Ingrosso2024}. 
In this work, however, I will mainly focus on neural coding and adaptation~\cite{Reinagel2001,Nikitin2009} to illustrate the results, although they are more broadly applicable. There, an input stimulus $X$ is encoded into an output observable $Y$ by a (group of) neuron(s)~\cite{Dayan2005}. 

Output events are generally sensitive to whether the input $X$ exceeds a given threshold; an event is considered ``on" when above this threshold, and ``off" otherwise. However, a single threshold loses much of the information contained in an input signal~\cite{Bialek2005}, while a system that can incorporate multiple readout mechanisms is able to resolve more details by ``binning'' expression levels into a discrete variable $Z = g(X)$. In neurology, it has been shown that such deterministic discretisation of the input signal is essential for optimal neural population coding~\cite{Bethge2003,Nikitin2009,Shao2023}, by maximising the mutual information $I(X,Y)$ between stimulus and response~\cite{Cover2006}.

However, when the input signal is discretised by an encoder, it must effectively maximise the \emph{retained} mutual information $I(Z,Y)$, i.e., between the compressed variable $Z$ and the output observable $Y$, subject to constraints on encoding cost, quantified by the Shannon entropy $H(Z)$. In Ref.~\cite{e22010007}, the authors compute an optimal trade-off between $H(Z)$ and $I(Z,Y)$ for a discrete, binned representation $Z$ of $X$ and a binary variable $Y$, subject to the Markov constraint $Z\leftrightarrow X\leftrightarrow Y$, which is related to a non-convex generalisation of the Deterministic Information Bottleneck (DIB)~\cite{Strouse2017,Strouse2019}.

The Pareto-optimal trade-off between $H_c \equiv H(Z)$ and $I_c \equiv I(Z,Y)$ is then characterized by 
\begin{equation}
    \label{eq:optimisation}
    I_c(H_c) \equiv \sup_{g:H(g(X))\leq H_c}{I(g(X),Y)}\,.
\end{equation}
This formulation contrasts with the usual scalarised IB approach, which minimizes a Lagrangian cost function $\mathcal{L} = \alpha H(Z) - (1-\alpha) I(Z,Y)$, and therefore finds only the concave hull of the Pareto front; intrusions are essentially ``skipped", leading to first-order phase transitions in the optimal solutions when tracing the front by varying $\alpha$~\cite{Rose1990}. Similar transitions in Pareto optimality can be found in boundary-driven morphogenesis~\cite{Berx_2025}, complex networks~\cite{Seoane2015}, learnability in feature learning~\cite{Wu2019}, biochemical discrimination~\cite{Berx_2024} or work fluctuations in stochastic thermodynamics~\cite{Solon2018,Forão_2025}, among others.

In this work, I apply the generalised DIB framework~\eqref{eq:optimisation} to a prototypical system in which a binary output variable of interest, $Y \in \{0,1\}$, is coupled to a slow stochastic input process $X$ that is compressed into a discrete representation $Z$. The process $X$ modulates the transition dynamics between the two states of $Y$ by shifting the underlying quasi-potential $\Delta G(X)$, with the values of $Y$ corresponding to distinct minima. 

This general setup can be mapped to different systems, ranging from biology to electrical engineering. The slowly varying input $X$ can for instance be a ligand concentration, sensory stimulus, environmental stress or an electrical signal in analog-to-digital converters (ADC)~\cite{Weaver2010}, while $Y$ is a fast, stochastic microscopic readout whose occupancy encodes instantaneous evidence about $X$; for example the phosphorylation state of a signalling protein or receptor~\cite{Markevich2004,Salazar2009,Amaral2017}, a short-time spike-rate regime in a neuron~\cite{Jones2015,Panzeri2022}, the instantaneous phenotypic state of a cell in an epigenetic landscape~\cite{Waddington2014,Moris2016}, or simply a digital signal. $Z$ is a finite-alphabet decision variable that bins the estimator of $P(Y|X)$ into a small number of reliably distinguishable output states, implemented biologically by distinct effectors or receptor expression profiles~\cite{Pushkar2015}, spike-count thresholds in neural coding~\cite{Chalk2018,Nikitin2009,Shao2023}, or population fractions, respectively, or digitally using unary coding~\cite{Yoffe2016}.

\begin{figure}[htp]
    \centering
    \includegraphics[width=0.9\linewidth]{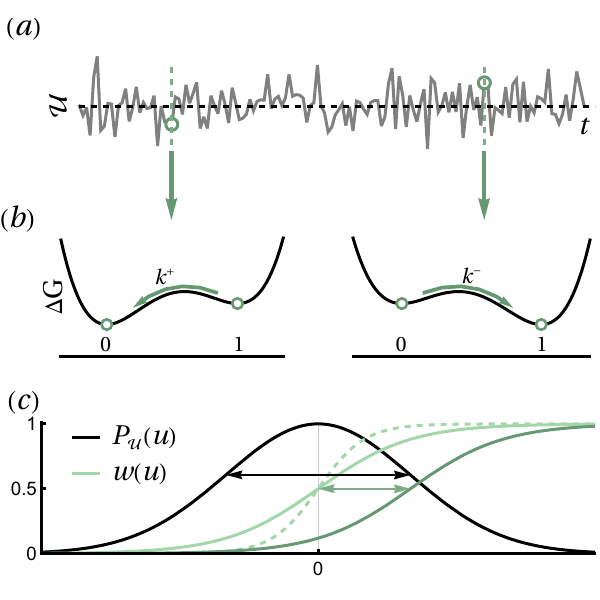}
    \caption{{\bf (a)} Single realisation of the steady-state Gaussian input signal fluctuations $\mathcal{U}$ as a function of time. Dashed vertical lines indicate possible scenarios in which the fluctuation is either negative (left) or positive (right) with respect to the steady-state mean (dashed black line). {\bf (b)} These input signals modulate transition rates $k^\pm$ between the states $Y\in\{0,1\}$ in a quasi-potential landscape $\Delta G$. {\bf (c)} The Gaussian fluctuations $P_\mathcal{U}(u)$ (black, scaled by $\sqrt{2\pi}$) superimposed on the response curve $w(u)$ (light green). Steeper (dashed green) and shifted (dark green) response functions are also shown.}
    \label{fig:plotgrid}
\end{figure}

To set the stage, assume $X$ can be described by the following Langevin equation
\begin{equation}
    \label{eq:langevin}
    \frac{dX}{dt} = -\frac{X-X^*}{\tau} + \eta(t),
\end{equation}
with adaptation timescale $\tau$ that drives a relaxation towards $X^*$, and $\eta$ a Gaussian white noise with mean zero, i.e., $\langle\eta(t)\rangle = 0$, and correlation $\langle \eta(t)\eta(t')\rangle = D \delta(t-t')$. The fluctuations with respect to the steady-state average $\langle X\rangle = X^*$, i.e., $\delta X = X-\langle X\rangle$ are Gaussian, possessing a stationary distribution $P_{\delta X}$ with variance $\sigma^2 = D\tau/2$. $X$ is subsequently coupled to a dynamic two-state system where a binary variable $Y$ switches between $Y=0$ and $Y=1$ with rates $k^\pm(X)$ that depend on the instantaneous value of $X$, assuming that relaxation of $Y$ is much faster than the slow dynamics of $X$. A cartoon representation of the system is drawn in Fig.~\ref{fig:plotgrid}(a,b).

The transition rates can be parametrised as $k^+ = \omega_0 \exp{(-\beta \Delta G_0)}$ and $k^- = \omega_0 \exp{(-\beta \Delta G_1)}$, with $\Delta G_{0,1}$ quasi-potential barriers that explicitly depend on $X$, $\beta = 1/k_{\rm B}T$ the inverse thermal energy and $\omega_0$ the switching attempt frequency. The probability for the two-state system to be in state $i$ at a time $t$ for a given $X$ is given by $w_i(t) = P(Y=i,t|X)$, $i=0,1$. This variable $W$ is a sufficient statistic for $X$, i.e., $I(W,Y) = I(X,Y)$. Eliminating $w_0(t)$ through $w_0(t) = 1-w_1(t)$ and dropping the subscript, the dynamics of $Y$ is then fully described by the master equation
\begin{equation}
    \label{eq:master_equation}
    \frac{\mathrm{d}w(t)}{\mathrm{d}t} = k^- w(t) - k^+ (1-w(t))\,,
\end{equation}
which, in steady state, yields
\begin{equation}
    \label{eq:PCW_general}
    w = \frac{k^+}{k^+ + k^-} = \frac{1}{1+e^{-\beta(\Delta G_1-\Delta G_0)}}\,.
\end{equation}
Assuming that the fluctuations of $X$ are much smaller than its mean, i.e., $\delta X\ll \langle X\rangle$. the quasi-potential differences are expanded to linear order, i.e.,
    \begin{equation}
        \label{eq:expansion}
        \Delta G_{0,1}(X) = \Delta G_{0,1}(\langle X\rangle) + \left.\frac{\mathrm{d}\Delta G_{0,1}}{\mathrm{d}X}\right|_{X=\langle X\rangle} \delta X + \mathcal{O}(\delta X^2)\,,
    \end{equation}
Rescaling the fluctuations $\mathcal{U} = \delta X/\sigma$ and setting $\kappa = \beta (\Delta G_1 - \Delta G_0)|_{u=0}$ as the bias and $\lambda = \beta \frac{d}{du}(\Delta G_1 - \Delta G_0)|_{u=0}$ as the sensitivity simplifies the steady-state response curve~\eqref{eq:PCW_general} to the general sigmoidal form
    \begin{equation}
        \label{eq:response_curve}
        w(u) = \frac{1}{1+\me^{-(\kappa + \lambda u)}}\,.
    \end{equation}
The steady-state Gaussian prior obtained by solving~\eqref{eq:langevin}, along with representative response curves~\eqref{eq:response_curve} are drawn in Fig.~\ref{fig:plotgrid}(c).

Shifting to the information-theoretical framework, $H(Z)$ and $I(Z,Y)$ can be computed, which are measured in bits. The Shannon entropy of the compressed variable $Z$ is defined as 
\begin{equation}
    \label{eq:shannon_Z}
    H(Z) = - \sum_{k=1}^M P_Z(k) \log_2{P_Z(k)}\,,
\end{equation}
where $M$ is the number of discretised states in the compressed representation $Z$, i.e., the number of bins. Similarly, the mutual information $I(Z,Y)$ is given by
\begin{equation}
    \label{eq:MI_ZY}
    I(Z,Y) = \sum_{k=1}^M \sum_{y=0,1} P_{Z,Y}(k,y) \log_2{\frac{P_{Z,Y}(k,y)}{P_Z(k)P_Y(y)}}\,.
\end{equation}
Thus, the key object required to compute the Pareto-optimal trade-off is the joint probability $P_{Z,Y}$, from which all marginals can be computed directly. Since $Z$ is defined by $W$ through $Z = k \Leftrightarrow W\in i_k$, with $i_k = [b_k, b_{k+1}]$ the $k$th bin of $W$, $P_{Z,Y}(k,y)$ is computed as
\begin{equation}
    \label{eq:P(Z,Y)}
    P_{Z,Y}(k,y) = \int_{i_k} [w_1(u)]^y \left[1-w_1(u)\right]^{1-y} P_{\mathcal{U}}(u)\mathrm{d}u\,,
\end{equation}
given a vector of bin edges $\mathbf{b} = (b_1,b_2,\dots,b_{M+1})$. Note that I use the same $i_k$ notation to denote bin intervals in $W$ and $\mathcal{U}$; I assume no confusion can arise due to the monotonic mapping~\eqref{eq:response_curve} between $W$ and $\mathcal{U}$. 

For a given $\mathbf{b}$, plugging~\eqref{eq:P(Z,Y)} and the marginals $P_Z,\,P_Y$ into~\eqref{eq:shannon_Z} and~\eqref{eq:MI_ZY} yields a single point in the $(H,I)$ phase space. By construction $b_1 = 0$ and $b_{M+1}=1$ such that there are $M-1$ internal bin edges, which constitute the degrees of freedom of the optimisation problem~\eqref{eq:optimisation}.

\begin{figure*}[htp]
    \centering
    \includegraphics[width=0.95\linewidth]{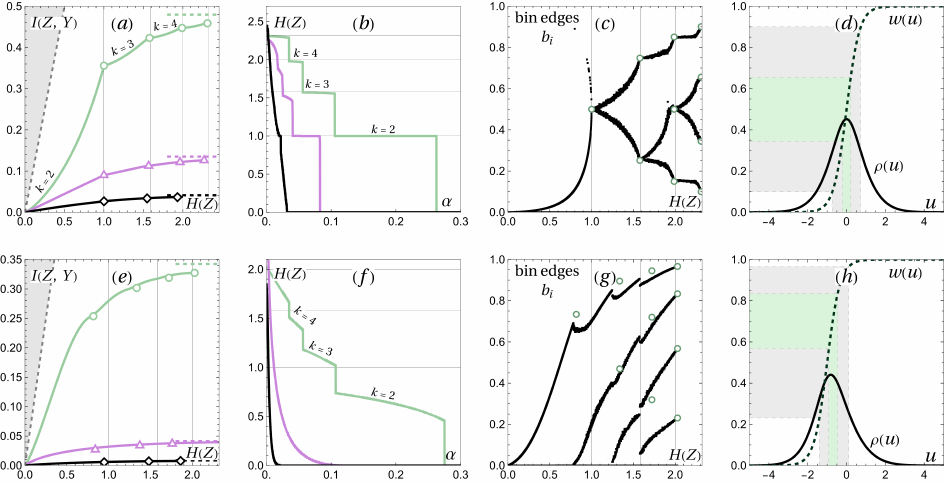}
    \caption{Pareto-optimal trade-offs for optimal binning with Gaussian prior for $\kappa = 0$ (a-d) and $\kappa=3$ (e-h), varying the sensitivity $\lambda$ (green: $\lambda =3$, purple: $\lambda =1$, black: $\lambda =1/2$). {\bf(a, e)} Pareto fronts show that for $\kappa=0$ and $\lambda > 1$, the optimal trade-off features sharp `corners' (open circles) indicating stable encoding choices. These corners seem absent when $\lambda \to 0$ or for $|\kappa|>0$. The maximum information $I(X,Y)$ is shown by the coloured dashed lines, with the gray region representing the unachievable bound $I(Z,Y) = H(Z)$. {\bf(b, f)} The convex hull of the fronts traces phase transitions between optimal bin numbers $k$, which may correspond to non-uniform bin edges. {\bf(c, g)} Optimal bin edges (with open circles from Lloyd's algorithm) and {\bf(d, h)} the resulting bin allocation for $M=5$ (shaded regions) are displayed for $\lambda=3$. Optimal allocation is compared with the predicted edge density~\eqref{eq:scaling_law} (full lines) and response curves~\eqref{eq:response_curve} (dashed lines). Vertical gridlines in (a, c, e, g) and horizontal ones in (b, f) denote the values of $H_k = \log_2{k}$.
    }
    \label{fig:paretoGaussian}
\end{figure*}

I use a NSGA-II evolutionary algorithm~\cite{Deb2008} to compute the Pareto-optimal trade-offs~\eqref{eq:optimisation} with high precision, given a choice of $(\kappa,\lambda)$. These Pareto fronts are shown in Fig.~\ref{fig:paretoGaussian}(a, e) for unbiased $(\kappa=0)$ and biased $(\kappa=3)$ systems, respectively. For zero bias, the fronts undergo a transition from profiles composed of piecewise convex segments separated by singular corner points for $\lambda \gtrsim 1$, to concave profiles for $\lambda \lesssim 1$. In the biased case, these singularities are smoothed out; however, for large $\lambda$, convex intrusions into the concave front persist.

For the piecewise trade-offs at large $\lambda$, each convex branch corresponds to a fixed number of bins $k\leq M$, where the convexity arises due to the local optimality of non-uniform bin sizes. At each corner point, a new bin is nucleated leading to an accelerating marginal information gain, albeit with diminishing returns. Creating an extra bin corresponds to acquiring an extra distinguishable readout state -- e.g. a new phosphorylation level, an extra firing-rate threshold in a postsynaptic neuron, a new population phenotype, or an additional comparator in an ADC. Between corners, the marginal benefit from creating a new readout state is negligible until pushed past a corner threshold — once past, the benefit jumps and a new readout state is created. The corner points constitute stable states, with first-order phase transitions between them, as shown in Fig.~\ref{fig:paretoGaussian}(b, f). Such phase transitions mirror predictions from neural population coding, where subpopulations of neurons are sequentially recruited to establish additional thresholds~\cite{Nikitin2009,Gjorgjieva2019,Shao2023}. This is consistent with neurological observations: nerve fibres connected to inner hair cells, which transduce sound into electrical signals, are organised into two or three subpopulations according to their sound-level thresholds~\cite{Liberman1982,Jackson2005}. From a scalarised DIB perspective, solutions lying on the convex branches are considered metastable; these metastability regions predict hysteresis: gaining a costly internal state may be harder than losing it, or vice versa.

The situation is markedly different for low $\lambda$ in Fig.~\ref{fig:paretoGaussian}(a, e); the Pareto front becomes globally concave, first-order transitions disappear for $k>2$, and the DIB~\eqref{eq:optimisation} coincides with its scalarised formulation. A system can gradually follow the Pareto front through continuous fine-tuning, i.e., by smoothly shifting threshold positions instead of creating entirely new readout states. This is similar to the phase transition from discrete to analog neural coding~\cite{Bethge2003}, where in the latter existing thresholds fine-tune through small continuous adjustments.

In these systems, no bin switching occurs any more; a trade-off with a number of bins $k_1$ strictly dominates one with $k_2 < k_1$. This resembles the IB in optimal sensing~\cite{Bauer2021}, where thresholded sensors-- corresponding to deterministic endpoints of IB with a fixed number of resolvable levels-- are almost on the optimal bounding curve.

Open symbols in Fig.~\ref{fig:paretoGaussian}(a, e) indicate the maximal points $(H,I)$ obtained from individual optimisations with fixed but increasing $M$. For cases exhibiting corners these points lie on the Pareto front and are therefore optimal, see Fig.~\ref{fig:paretoGaussian}(a). Conversely, when a system with $M$ bins is allowed to nucleate an additional bin, this previously optimal point may fall below the new Pareto front, becoming suboptimal, see Fig.~\ref{fig:paretoGaussian}(e), or enter a convex intrusion, corresponding to a metastable configuration accessible only through a hysteretic adjustment of the bin edges.

Gridlines in Fig.~\ref{fig:paretoGaussian} denote the maximum entropy for $k$ equiprobable bins, $H_k = \log_2{k}$, while the respective local maxima for a \emph{fixed} number of bins---denoted by open symbols in panels (a) and (e)--- are generally located at entropies slightly lower than $H_k$. Thus, non-equiprobable bins constitute a more efficient strategy; it is better to allocate more resolution to informative regions of the signal than to use uniform bin spacing. When constructing the binned partition, the system is effectively representing each posterior range of values by its centroid and minimising the expected Kullback-Leibler (KL) divergence between the two. As such, an optimal internal bin edge sits where the KL divergence to the two neighbouring centroids is equal. Given an initial partition, the centroids can be computed, from which a new partition can be derived. Iterating this procedure until convergence corresponds to Lloyd's algorithm~\cite{Lloyd1982}, used, e.g., for K-means clustering in machine learning, but using the KL divergence as the distortion measure (see SM~\cite{SM}). 

Asymptotically, for a large but fixed number of bins $M$, the point density of bin edges $\rho(u)$ scales as 
\begin{equation}
    \label{eq:scaling_law}
    \rho(u) \propto \left[P_\mathcal{U}(u)\mathcal{I}_F(u)\right]^\frac{1}{3}\,,
\end{equation}
where $\mathcal{I}_F(u)$ is the Fisher information about $u$ for the encoder family $P(Y|\,U=u)$. This cube-root scaling constitutes the central result of this work, and can be derived using high-rate quantisation theory, see SM~\cite{SM}. $\mathcal{I}_F(u)$ is computed exactly using~\eqref{eq:response_curve}, i.e., $\mathcal{I}_F(u) = (\lambda^2/4)\sech^2\{(\kappa + \lambda u)/2\}$. Intuitively, the density scaling means that doubling the informativeness of a region, quantified by $\mathcal{I}_F(u)$, does not double the number of thresholds there -- it only increases them by the cube root. Evolving one extra bin thus leads to progressively smaller informational payoff when many thresholds are already present; this could explain why many biological systems generally use only a small number of qualitatively different internal states. 

In Fig.~\ref{fig:paretoGaussian}(c, g), the Pareto-optimal bin edges (black) are shown as a function of $H(Z)$ for $\lambda=3$; green open symbols denote the optimal bin edges for fixed $M$, independently computed by Lloyd's algorithm. Note that in panel (c) they align perfectly with the points where a new bin is nucleated in the Pareto front, indicating that all bin edges that locally maximize $I(Z,Y)$ are indeed given by Lloyd's algorithm when the front is piecewise convex. When the front shows concavity, however, the same argument as before holds: optimal bin edges for a fixed $M$ can become suboptimal when that $M$ is increased. Lloyd's algorithm then only gives the exact bin edges for $k = M$ (and not for $k<M$), which are indicated by the rightmost set of open circles in panel (g).

In Fig.~\ref{fig:paretoGaussian}(d, h), the optimal bin edges from panels (c, g) are mapped back to $u-$space. The resulting bins (shaded regions) show that the edges are indeed clustered around the most informative region, determined by the maximum of $\rho(u)$. The scaling~\eqref{eq:scaling_law} seems to hold qualitatively already for low $M$. 

To go beyond the Gaussian assumption, let us shift to a noise source that can produce non-Gaussian fluctuations. Assume that the system experiences delta-function `kicks' at random times $\{t_k\}$, which are Poisson distributed with rate $\nu$, and where the kick amplitudes $A_k$ are i.i.d. random variables with zero-mean distribution $P_A(a)$, i.e., in ~\eqref{eq:langevin} the noise is $\eta(t) = \sum_k A_k \delta(t-t_k)$ with $\langle A\rangle=0$. This is an instance of so-called \emph{shot noise}. The characteristic function (CF) for the fluctuations is derived in the SM~\cite{SM} and depends on the ability to compute the CF for the amplitude distribution. An illustrative yet analytically tractable choice for $P_A$ is the Laplace distribution $P_A(a) = \frac{1}{2b}\exp{\{-|a|/b\}}$, with $b > 0$ the typical size of one kick. The rescaled prior $P_\mathcal{U}(u)$ is given by 
\begin{equation}
    \label{eq:PU_nonGaussian}
    P_{\mathcal{U}}(u) = C(\mu) |u|^{\mu-\frac{1}{2}} K_{\mu-\frac{1}{2}}(\sqrt{2\mu}\,|u|)\,,
\end{equation}
with $\mu = \nu \tau/2$ the control parameter for non-Gaussianity (the excess kurtosis is $\gamma_2 = 3/\mu$). $K_{n}(x)$ is the modified Bessel function of the second kind~\cite{abramowitz} and $C(\mu)$ is a normalisation factor. Equation~\eqref{eq:PU_nonGaussian} reduces to well-known functions for particular choices of $\mu$. For instance, for $\mu = 1$, it becomes the Laplace distribution with scale parameter $1/\sqrt{2}$, while for $\mu\rightarrow\infty$ the Gaussian prior is recovered.

\begin{figure}[tp]
    \centering
    \includegraphics[width=\linewidth]{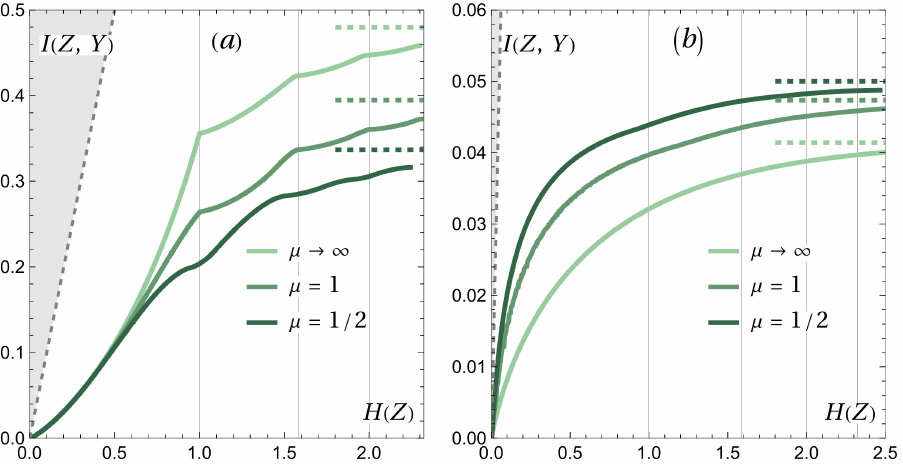}
    \caption{Information-cost Pareto fronts for non-Gaussian prior, with {\bf(a)} $\lambda=3,\,\kappa=0$ and {\bf(b)} $\lambda=1,\,\kappa=3$. For $\kappa=0$ the Gaussian limit results in a globally more optimal trade-off. Conversely, higher $|\kappa|$ can result in finite $\mu$ becoming the optimal curve.
    }
    \label{fig:pareto_nonGaussian}
\end{figure}

Proceeding in the same fashion as for the Gaussian case, the Pareto front in $(H,I)-$space can be computed. In Fig.~\ref{fig:pareto_nonGaussian}, the Pareto fronts for two cases are shown: $\mu=1$ (Laplace distribution) and $\mu = 1/2$ (product-normal distribution), plotted together with those obtained in the Gaussian limit. As before, unbiased systems with $\kappa=0$ exhibit piecewise convex fronts. Increasing $\mu$ leads to progressively more favourable global trade-offs, whereas fronts corresponding to smaller $\mu$ are entirely dominated by those with larger $\mu$. 

\begin{figure}[htp]
    \centering
    \includegraphics[width=\linewidth]{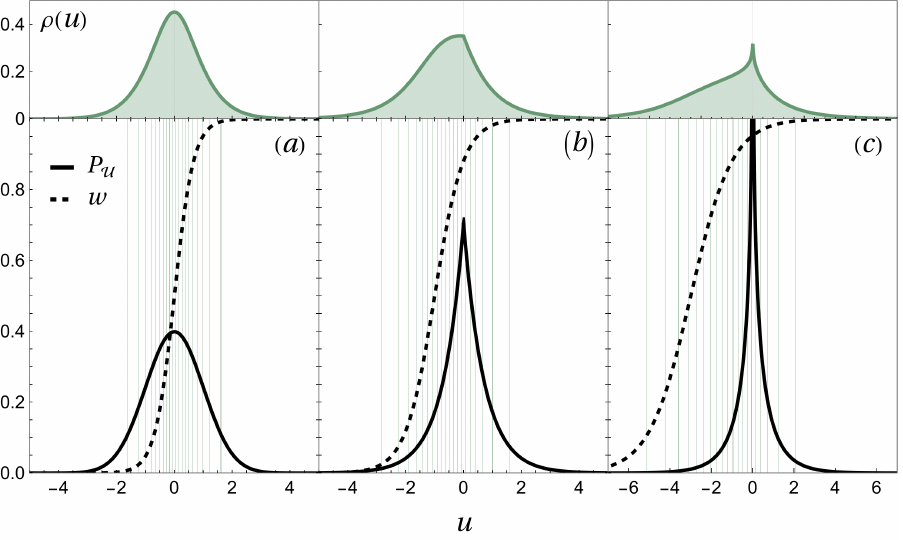}
    \caption{Numerically computed Pareto-optimal distribution of bin edges (thin vertical lines) for $M=20$, superimposed on the priors $P_\mathcal{U}$ (full lines) and response functions $w(u)$ (dashed lines). The top row of figures shows the corresponding bin edge density $\rho(u)$~\eqref{eq:scaling_law}, which agrees well with the numerical results. {\bf(a)} Gaussian prior $\mu\rightarrow\infty$ with $\lambda=3,\,\kappa=0$; {\bf(b)} Laplacian prior $\mu=1$ with $\lambda=\kappa=2$; {\bf(c)} product-normal prior $\mu=1/2$ with $\lambda=1,\,\kappa=3$.}
    \label{fig:Fisher_nonGaussian}
\end{figure}

For sufficiently small $\lambda$ and high $|\kappa|$, however, this behaviour changes, as illustrated in Fig.~\ref{fig:pareto_nonGaussian}(b). Decreasing $\mu$ in~\eqref{eq:PU_nonGaussian} shifts a larger fraction of the probability mass into the tails compared to the Gaussian case. Depending on the degree of non-Gaussianity of the input noise, bin edges are shifted accordingly to better cover the most informative regions of the input. This qualitative effect is shown in Fig.~\ref{fig:Fisher_nonGaussian} by substituting equation~\eqref{eq:PU_nonGaussian} into the main result~\eqref{eq:scaling_law}. If the Pareto front for a non-Gaussian prior globally dominates that for the Gaussian surrogate, the same level of mutual information can be achieved at a lower encoding cost, either by using a lower number of discrete states or shifting existing thresholds in a non-uniform manner. Note that this conclusion holds for \emph{optimised} encoders: a fixed, non-adapted readout may not realise the advantage. As a result, sensitivity to model mismatch can be assessed experimentally, for instance by measuring performance of encoders optimised for Gaussian inputs when exposed to non-Gaussian signals.

The optimal bin edges for large $\lambda$ colocate near the region where $w(u)$ changes drastically, i.e., near $u^* = -\kappa/\lambda$, but are also influenced heavily by the prior distribution, see Fig.~\ref{fig:Fisher_nonGaussian}, where e.g., in panel (a) bin edges cluster around $u = 0$, while for panel (b) they cluster near $u\approx -1$, but their distribution is skewed towards lower values of $u$. For small $\lambda$, the maximum of $\rho(u)$ is determined approximately by that of $P_\mathcal{U}(u)$, but most probability mass is still located near $u^*$, see panel (c).

In conclusion, I characterised the geometry of Pareto-optimal lossy compression of a fluctuating signal $X$ into a finite discrete readout $Z$ that preserves information about a binary output variable $Y$. Combining Pareto optimisation with asymptotic quantisation theory, I show that the information–cost frontier is generically nontrivial: in the small-bias or high-sensitivity regime it fragments into piecewise-convex branches with sharp corners where new readout states nucleate, while in the opposite regime the frontier is smoothly concave and can be traced by continuously tuning thresholds. In the high-rate limit the local threshold density follows a universal cube-root law~\eqref{eq:scaling_law}, linking the prior and Fisher information and explaining why only a few well-placed thresholds are typically optimal. Extending to non-Gaussian priors reveals that heavy tails can markedly alter optimality --- a non-Gaussian environment can be cheaper to encode, for the same retained information, than a Gaussian one --- so sensory systems adapted to such environments may need fewer readout states or lower cost than Gaussian theory predicts. These results unify ideas from information bottlenecks, high-rate quantisation theory, and biological sensing, and make concrete, testable predictions, e.g., hysteresis or abrupt state acquisition, for engineered ADCs and molecular or neural encoders. With straightforward extensions, e.g., categorical $Y$ or non-equilibrium dissipation~\cite{Mehta2012,Wolpert2024}, this approach points toward a unified, predictive theory of lossy compression and information flow in biological systems.

\begin{acknowledgments}
The author thanks K. Proesmans for enlightening discussions on the topic. This research is supported by the Novo Nordisk Foundation with grant No. NNF18SA0035142.
\end{acknowledgments}

\bibliography{biblio}

\end{document}


\author{Jonas Berx}
\affiliation{Niels Bohr International Academy, Niels Bohr Institute, University of Copenhagen, Blegdamsvej 17, 2100 Copenhagen, Denmark}
\title{Supplemental Material: Geometry and universal scaling of Pareto-optimal signal compression}
\date{\today}

\maketitle

This Supplemental Material contains detailed mathematical proofs on (i) the optimal bin partition for solutions on the information-cost Pareto front, showing that it reduces to Lloyd's algorithm with the Kullback-Leibler divergence as distortion measure; (ii) the scaling behaviour of the bin density in the $M\rightarrow\infty$ limit, and (iii) the derivation of the non-Gaussian shot-noise input prior.

\section{Proof of the optimal bin partition}\label{sec:partition}
Given the distribution $P_\mathcal{U}(u)$ supported on the real line and the posterior conditional distribution vector $Q(u) = \left(Q_y(u)\right)_{y\in\mathcal{Y}}$, with $Q_y(u) = P(Y=y |\, \mathcal{U} = u)$, with finite or countable support $\mathcal{Y}$, I show that the optimal partition of $u-$space is given by the Lloyd algorithm~\cite{Lloyd1982} using the Kullback-Leibler divergence as a distortion measure. Instead of partitioning $u-$space directly, I bin $W-$space instead, using the posterior distribution to map back if necessary.

Let us denote an admissible partition into $M$ bins by the vector $\mathbf{b}$ of its bin edges, i.e., $\mathbf{b} = [b_1,b_2,\dots,b_{M+1}]$. By construction, one can always take $b_1 = 0$ and $b_{M+1} = 1$, and I denote bin $k$ as $i_k = [b_k,b_{k+1}]$. The optimal partition can be found by varying internal edges $b_j$ for $j=2,\dots,M$. 

For bin $k$, define the bin probability and the bin-averaged posterior respectively as
\begin{equation}
    \label{eq:bin_prob}
    P_Z(k) = \int_{i_k}P_\mathcal{U}(u)\mathrm{d}u\,,
\end{equation}
and
\begin{equation}
    \label{eq:bin_average}
    \theta_k = P(Y | \,Z=k) = \frac{1}{P_Z(k)} \int_{i_k} P_\mathcal{U}(u) Q(u) \mathrm{d}u\,,
\end{equation}
where $\theta_k$ is a probability vector on $\mathcal{Y}$, i.e., $\sum_{y\in\mathcal{Y}} \theta_{k,y} = 1$. 

Maximising $I(Z,Y)$ for a given number of bins $M$ is identical to minimising the conditional entropy $H(Y|Z)$, since in the definition of $I(Z,Y) = H(Y) - H(Y|Z)$ the term $H(Y)$ does not depend on the partition. 

It can be easily shown that $H(Y|Z)$ can be decomposed as follows
\begin{equation}
    \label{HYZ_decomposition}
    H(Y|Z) = H(Y|U) + \sum_{k=1}^M\int_{i_k} P_\mathcal{U}(u)D_{KL}\left(Q(u)||\theta_k\right)\mathrm{d}u\,,
\end{equation}
where $D_{KL}(Q(u)||\theta_k) = \sum_y Q_y(u) \log_2\left( Q_y(u)/\theta_{k,y}\right)$ is the Kullback-Leibler divergence between the posterior and its bin-averaged representation. Once again, $H(Y|U)$ is independent of the exact partition such that a minimisation of $H(Y|Z)$ reduces to minimising the second term in the decomposition~\eqref{HYZ_decomposition}. We now find necessary conditions on the bin edges $\{b_j\}_{j=2}^M$ and the bin-averaged posteriors $\{\theta_k\}_{k=1}^M$.

Assuming that the bin edges are fixed and varying $\theta_k$ under the constraint $\sum_y \theta_{k,y} = 1$ with $\theta_{k,y} > 0$, the functional
\begin{equation}
    \mathcal{J}(\{\theta_k\}) = -\sum_{k=1}^M\int_{i_k} P_\mathcal{U}(u)\sum_{y}Q_y(u) \log_2\theta_{k,y}\mathrm{d}u\,,
\end{equation}
can be minimised, where terms that do not involve $\theta_k$ have been discarded. For each $k$ the minimisation to be performed is then 
\begin{equation}
    \min_{\theta_k:\sum_y\theta_{k,y}=1} \left(-\int_{i_k} P_\mathcal{U}(u)\sum_{y}Q_y(u) \log_2\theta_{k,y}\mathrm{d}u\right)\,.
\end{equation}

The functionals for optimisation can be set as
\begin{equation}
    \mathcal{L}(\theta_k,\mu_k) = -\int_{i_k} P_\mathcal{U}(u)\sum_{y}Q_y(u) \log_2\theta_{k,y}\mathrm{d}u + \mu_k \left(\sum_y \theta_{k,y} -1\right)\,,
\end{equation}
using a Lagrange multiplier $\mu_k$ to ensure normalisation. Differentiating with respect to $\theta_{k,y}$ and setting the result to zero yields, for all $y$:
\begin{equation}
    \label{eq:theta_ky}
    \theta_{k,y} = \frac{1}{P_Z(k)} \int_{i_k} P_\mathcal{U}(u) Q_y(u)\mathrm{d}u\,,
\end{equation}
where the normalisation condition yields $\mu_k = P_Z(k)$. Equation~\eqref{eq:theta_ky} is the first-order stationarity condition for minimizing $\mathcal{J}$ over $\theta_k$ and, as such, any first-order variation of $H(Y|Z)$ arising from infinitesimal changes of $\theta_k$ will vanish when this equation holds.

Consider now the variation of a single internal bin edge $b_j$ for some $j$ with $2\leq j\leq M$. This edge separates bins $k = j-1$ and $k = j$. Let $\mathcal{H}(\mathbf{b},\mathbf{\theta})$ denote the objective $H(Y|Z)$ written as a function of the vector of bin edges and the centroids $\theta = (\theta_1,\dots,\theta_M)$, i.e.,
\begin{equation}
    \label{eq:H_btheta}
    \mathcal{H}(\mathbf{b},\theta) = H(Y|U) + \sum_{k=1}^M \int_{b_k}^{b_{k+1}} P_\mathcal{U}(u) D_{KL}\left(Q(u)||\theta_k\right)\mathrm{d}u\,,
\end{equation}
the derivative of $\mathcal{H}$ with respect to $b_j$ is computed as
\begin{equation}
    \label{eq:H_derivative}
    \frac{\mathrm{d}\mathcal{H}(\mathbf{b},\theta(\mathbf{b}))}{\mathrm{d}b_j} = \left.\frac{\partial\mathcal{H}}{\partial b_j}\right|_{\theta \, \text{fixed}} + \sum_{k=1}^M \sum_y \frac{\partial\mathcal{H}}{\partial \theta_{k,y}}\frac{\partial\theta_{k,y}}{\partial b_j}\,.
\end{equation}
There is thus a direct effect of moving the integration limits and an indirect effect via how the centroids $\theta_k$ change with the boundaries. To compute the first term, note that only two of the integrals in the sum in equation~\eqref{eq:H_btheta} explicitly depend on $b_j$: the ones for $k=j-1$ and $k=j$. Hence, the first term of equation~\eqref{eq:H_derivative} can be reduced to only two terms, i.e.,
\begin{equation}
    \left.\frac{\partial\mathcal{H}}{\partial b_j}\right|_{\theta \, \text{fixed}} = P_\mathcal{U}(b_j) \left[D_{KL}\left(Q(b_j)||\theta_{j-1}\right) - D_{KL}\left(Q(b_j)||\theta_{j}\right)\right]\,.
\end{equation}
Moving the boundary infinitesimally takes probability mass at $b_j$ out of bin $j-1$ and transfers it to bin $j$; the net change in the objective density is the difference of the local KL penalties weighted by the input density at that point.

To compute the second term in~\eqref{eq:H_derivative}, the derivative $\partial \mathcal{H}/\partial \theta_{k,y}$ is essential. Recall, however, from the first-order stationary condition that at the centroid-optimal $\theta_k$, the condition $\partial \mathcal{H}/\partial \theta_{k,y} = 0$ holds for all $k,\,y$, subject to the normalisation constraint. Because $\theta$ were chosen to minimize $\mathcal{H}$ for the given partition, the directional derivative of $\mathcal{H}$ in any feasible direction of $\theta$ is zero; hence the inner product in equation~\eqref{eq:H_derivative} above vanishes to first order.

A necessary condition for a stationary partition is thus given by
\begin{equation}
    \frac{\mathrm{d}\mathcal{H}(\mathbf{b},\theta(\mathbf{b})}{\mathrm{d}b_j} = P_\mathcal{U}(b_j) \left[D_{KL}\left(Q(b_j)||\theta_{j-1}\right) - D_{KL}\left(Q(b_j)||\theta_{j}\right)\right] = 0\,,
\end{equation}
yielding the final \emph{equidistance} condition on the centroids, i.e.,
\begin{equation}
    \label{eq:equidistant}
    D_{KL}\left(Q(b_j)||\theta_{j-1}\right) = D_{KL}\left(Q(b_j)||\theta_{j}\right)\,, \qquad \mbox{\text{for each $j=2,\dots,M$.}}
\end{equation}
The set of equations~\eqref{eq:bin_average} and~\eqref{eq:equidistant} form a coupled fixed-point system which can be iterated to find the optimal partition into $M$ bins. This is exactly Lloyd's algorithm~\cite{Lloyd1982}, which is commonly used in machine learning to find centroidal Voronoi tesselations for K-means data clustering schemes.

\section{The large bin number scaling limit}
Assume that the input probability density $P_\mathcal{U}(u)$ has compact support on a set $\mathcal{V}$ and is twice continuously differentiable, i.e., $P\in C^2$, and that the local posterior $Q(u)$ is is a smooth map into the probability simplex $\mathcal{Y}$. Additionally assume $Q\in C^3$ such that the Taylor series below are valid. The support is binned optimally according to Lloyd's algorithm into a large number $M$ of contiguous Voronoi bins $i_k = [b_k,b_{k+1}]$ with Voronoi centres $u_k$. Let $\Delta_k = b_{k+1} - b_k$ be the bin widths. Considering the optimal partition that minimises the expected KL distortion, i.e., the second term in equation~\eqref{eq:H_btheta}, I now derive the asymptotic density $\rho(u)$ of cell centres as $M\rightarrow\infty$.

Fix a bin $i_k$ with centroid $\theta_k$. Under the centroid condition the optimal choice for a small bin is $\theta_k \approx Q(u_k)$; for a point $u = u_k + x$, expand $Q(u)$ and the KL divergence around $Q(u_k)$, i.e.,
\begin{align}
    Q(u) &= Q_0 + Q'(u_k)x + \frac{1}{2}Q''(u_k)x^2 + \mathcal{O}(x^3) \label{eq:Q_expand} \\
    D_{KL}(Q(u)||Q_0) &= \frac{x^2}{2}Q'(u_k)^\intercal H(Q_0) Q'(u_k) + \mathcal{O}(|x|^3)\,, \label{eq:KL_expand}
\end{align}
where $Q_0 = Q(u_k)$ and where $H(Q_0) = (\partial^2 / \partial\theta^2) D_{KL}(Q||\theta)|_{Q=\theta=Q_0}$ is the Hessian of the KL divergence. Define the Fisher coefficient $\mathcal{I}_F(u) \equiv Q'(u)^\intercal H(Q(u)) Q'(u)$, such that for small $x$,
\begin{equation}
    D_{KL}(Q(u_k + x)||Q(u_k)) = \frac{x^2}{2}\mathcal{I}_F(u_k)\,.
\end{equation}

For the Voronoi tesselation in the large $M$ limit, it can be assumed in good approximation that each bin is symmetric around its center $u_k$, to leading order. The per-bin contribution to the second term of equation~\eqref{eq:H_btheta} is then
\begin{equation}
    \begin{split}
        H_k &= \int_{i_k} P_\mathcal{U}(u) D_{KL}(Q(u)||Q(u_k))\mathrm{d}u \\
        &= \int\limits_{-\Delta_k/2}^{\Delta_k/2} P_\mathcal{U}(u_k + x) \left(\frac{x^2}{2}\mathcal{I}_F(u_k) + \mathcal{O}(|x|^3)\right)\mathrm{d}x\,.
    \end{split}
\end{equation}
Expanding $P_\mathcal{U}(u_k + x) = P_\mathcal{U}(u_k) + \mathcal{O}(x)$, this can be inserted into the expression for $H_k$. Keeping the leading term, this yields
\begin{equation}
    H_k = \frac{1}{24}P_\mathcal{U}(u_k) \mathcal{I}_F(u_k)\Delta_k^3 + \mathcal{O}(\Delta_k^4)\,,
\end{equation}
where the factor $1/24$ comes from the symmetric integration of $x^2/2$. For the optimal partition, the total distortion that needs to be minimised is then equal to $H_{\rm dist} = \sum_k H_k$. Assuming that the partition is fine enough (i.e., $M$ is large), the summation can be approximated by an integral and the point density $\rho(u)$ can be defined, approximating $1/\Delta_k$ as $u\approx u_k$ in the continuum limit. There are $\rho(u)\mathrm{d}u$ bins in the interval $[u,u+\mathrm{d}u]$, each contributing $(1/24)P_\mathcal{U}(u) \mathcal{I}_F(u) \Delta(u)^3$. Consequently, the local distortion density is given by
\begin{equation}
    \mathrm{d}H_{\rm dist} \approx \frac{1}{24}P_\mathcal{U}(u) \mathcal{I}_F(u) \Delta(u)^3 \rho(u)\,\mathrm{d}u\,.
\end{equation}
The continuum approximation to the total distortion is therefore
\begin{equation}
    H_{\rm dist} \approx \frac{1}{24}\int_\mathcal{V} P_\mathcal{U}(u) \mathcal{I}_F(u) \rho^{-2}(u) \mathrm{d}u\,,
\end{equation}
with remainder terms that vanish as $\mathcal{O}(M^{-3})$ for $M\rightarrow\infty$. To solve the constrained minimisation of $H_{\rm dist}$ subject to the normalisation $\int_\mathcal{V}\rho(u)\mathrm{d}u = M$, the Lagrangian functional
\begin{equation}
    \mathcal{L}(\rho) = \int_\mathcal{V}\left[ \frac{P_\mathcal{U}(u) \mathcal{I}_F(u)}{24\, \rho^{2}(u)} + \mu \rho(u)\right]\mathrm{d}u\,,
\end{equation}
with Lagrange multiplier $\mu$ is defined. Taking the functional derivative and requiring pointwise stationarity, i.e.,
\begin{equation}
    \frac{\delta\mathcal{L}}{\delta\rho} = -\frac{P_\mathcal{U}(u) \mathcal{I}_F(u)}{12\, \rho^{3}(u)} + \mu = 0\,,
\end{equation}
I finally find that $\rho(u) \propto \left(P_\mathcal{U}(u)\mathcal{I}_F(u)\right)^{1/3}$, where the proportionality constant $c$ can be fixed through normalisation, i.e., $\int_{\mathcal{V}}\left(P_\mathcal{U}(s)\mathcal{I}_F(s)\right)^{1/3}\mathrm{d}s = M/c$.

\section{Derivation of the stationary PDF under shot noise}
Starting from equation~(2) in the main text,
\begin{equation}
    \label{eq:langevin}
    \frac{dX}{dt} = -\frac{X-X^*}{\tau} + \eta(t),
\end{equation}
assume that the system experiences delta-function `kicks' at random times $\{t_k\}$, which are Poisson distributed with rate $\nu$, and where the kick amplitudes $A_k$ are i.i.d. random variables with a mean-zero distribution $P_A(a)$, i.e., $\eta(t) = \sum_k A_k \delta(t-t_k)$. The zero mean condition is chosen here to guarantee that the process mean indeed evolves towards $X^*$; if it is non-zero the mean of the steady-state solution of~\eqref{eq:langevin} is given by $\langle X\rangle = X^* + \nu \tau \langle A\rangle$. The fluctuations $\delta X(t) = X(t) - \langle X\rangle$ thus follow 
\begin{equation}
    \label{eq:langevin_nonzero_mean}
    \frac{d\delta X(t)}{dt} = -\frac{\delta X(t)}{\tau} + \xi(t)\,,
\end{equation}
with $\xi(t) = \eta(t) - \nu \langle A\rangle$ a zero-mean centred shot noise. As such, $\langle A\rangle = 0$ is chosen without significant loss of generality. Between two consecutive kicks $t\in[t_0,t_1]$, the system evolution is deterministic, i.e.,
\begin{equation}
    \delta X(t) = \delta X(t_0) \me^{-(t-t_0)/\tau}\,.
\end{equation}

Assuming the process has started some time in the past $-T$, $T>0$ with $\delta X(-T) = \delta X_0$ the initial condition, the contribution each random kick makes can be propagated in time; individually each kick contributes $A_k \exp{\{-(t-t_k)/\tau\}}$ to $\delta X(t)$. Summing contributions and taking the initial condition into account, the fluctuations are given at time $t$ by
\begin{equation}
    \delta X(t) = \delta X_0 \me^{-(t+T)/\tau} + \sum_{-T<t_k<t} A_k \me^{-(t-t_k)/\tau}\,.
\end{equation}
Taking the limit $T\rightarrow\infty$, the first term vanishes and the general solution is given by $\delta X(t) = \sum_{t_k<t} A_k \me^{-(t-t_k)/\tau}$, where the sum now runs over the entire history. To find the stationary probability of this process, the characteristic function can be computed, which is given by
\begin{equation}
    \varphi_{\delta X}(u) = \left\langle\me^{i u \delta X}\right\rangle = \left\langle\exp{\left(i u \sum_{t_k<t} A_k \me^{-(t-t_k)/\tau}\right)}\right\rangle\,.
\end{equation}
To proceed, I use Campbell's theorem~\cite{Kingman1992}, such that expectation values of point processes can be written as integrals of the form
\begin{equation}
    \label{eq:campbell}
    \ln{\left\langle\me^{i u \sum_k f(t_k,A_k)}\right\rangle} = \nu \int_0^\infty \left(\left\langle\me^{iuf(s,A)}\right\rangle-1\right)\mathrm{d}s\,.
\end{equation}
Parametrising $s = t - t_k > 0$ and setting $f(s,A) = A \me^{-s/\tau}$, the characteristic function can be written as
\begin{equation}
    \label{eq:charfunX}
    \varphi_{\delta X}(u) = \exp{\left\{\nu\int_0^\infty \left(\left\langle\me^{i u A \me^{-s/\tau}}\right\rangle-1\right)\right\}} = \exp{\left\{\nu\int_0^\infty \left(\varphi_A(u \me^{-s\tau)})-1\right)\right\}}\,,
\end{equation}
such that computing $\varphi_{\delta X}$ boils down to computing the characteristic function $\varphi_A$ of the kick amplitudes.

To go beyond Gaussian fluctuations while maintaining analytical tractability, I choose the cenetered Laplace distribution $P_A(a) = \frac{1}{2b}\exp{\{-|a|/b\}}$, with $b > 0$ the typical size of one shot in the compound-Poisson input. The characteristic function for this process can easily be computed in closed form, 
\begin{equation}
    \varphi_A(v) = \int_{-\infty}^\infty P_A(a) \me^{i v a}\mathrm{d}a = \frac{1}{2b}\int_{-\infty}^\infty \me^{-\frac{|a|}{b}} \me^{i v a}\mathrm{d}a = \frac{1}{1+b^2 v^2}\,.
\end{equation}
Plugging this into equation~\eqref{eq:charfunX},
\begin{equation}
    \varphi_{\delta X}(u) = \exp{\left\{\nu \int_0^\infty \left[\frac{1}{1+u^2 b^2 \me^{-2 s/\tau}}-1\right]\mathrm{d}s\right\}} = (1+b^2 u^2)^{-\mu}\,,
\end{equation}
with $\mu = \nu\tau/2$. From $\varphi_{\delta X}$, the moments and cumulants of the distribution $P_{\delta X}$ can be computed. While the mean is identically zero, the variance is given by $\sigma^2 = 2 b^2 \mu$ and the excess kurtosis by $\gamma_2 = 3/\mu$. From the characteristic function, it can easily be seen that the Gaussian noise limit with unit variance is recovered by making the choice $b = 1/\sqrt{2\mu}$ and letting $\mu\rightarrow\infty$. The probability distribution of $\delta X$ can now be computed from $\varphi_{\delta X}(u)$ by Fourier transformation, i.e., 
\begin{equation}
    P_{\delta X}(x) = \frac{1}{2\pi} \int_{-\infty}^\infty \me^{-i u x}\varphi_{\delta X}(u)\mathrm{d}u = \frac{|x|^{\mu-\frac{1}{2}}}{2^{\mu-\frac{1}{2}} b^{\mu+\frac{1}{2}} \sqrt{\pi} \Gamma(\mu)} K_{\mu-\frac{1}{2}}\left(\frac{|x|}{b}\right)\,,
\end{equation}
where $\Gamma(\mu)$ is the gamma function and $K_n(x)$ is the modified Bessel function of the second kind~\cite{abramowitz}.

Finally, to match the form $P_\mathcal{U}(u)$ required for the information-theoretic framework, the fluctuations are rescaled by the standard deviation $\sigma$, i.e., define $\mathcal{U} = \delta X/\sigma$ such that
\begin{equation}
    \label{eq:non_gaussian_prior_rescaled}
    P_{\mathcal{U}}(u) = C(u) |u|^{\mu-\frac{1}{2}} K_{\mu-\frac{1}{2}}(\sqrt{2\mu}\,|u|)\,, \qquad \mbox{with} \qquad C(u) = \frac{\mu^{\frac{\mu}{2}+\frac{1}{4}}}{2^{\frac{\mu}{2}-\frac{3}{4}}\sqrt{\pi}\,\Gamma(\mu)}\,.
\end{equation}

\begin{figure}[htp]
    \centering
    \includegraphics[width=0.5\linewidth]{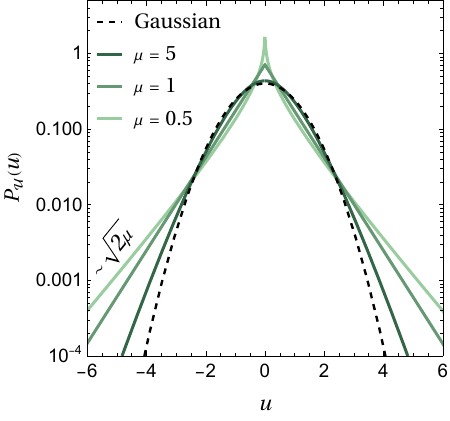}
    \caption{Log plot of the non-Gaussian prior distributions (coloured lines; full) and the Gaussian prior (dashed), showing the heavy-tailedness of eq.~\eqref{eq:non_gaussian_prior_rescaled}. The exponential scaling of the tails for finite $\mu$ is shown by the slope $\sqrt{2\mu}$.}
    \label{fig:non_gaussian_prior}
\end{figure}

For some particular values, the expression~\eqref{eq:non_gaussian_prior_rescaled} reduces to known functions, which are shown in Tbl.~\ref{tab:values}. The distribution $P_\mathcal{U}(u)$ is shown in Fig.~\ref{fig:non_gaussian_prior} for different $\mu$ and compared with a Gaussian prior.

\begin{table}[htp]
    \centering
    \begin{tabular}{c@{\hskip 0.15in}||@{\hskip 0.25in}c@{\hskip 0.25in}|@{\hskip 0.25in}c@{\hskip 0.25in}|@{\hskip 0.25in}c@{\hskip 0.25in}|@{\hskip 0.25in}c}
        $\mu $ & $0$ & $ 1/2$ & $1$ & $\infty$\\
        \hline
        $P_\mathcal{U}(u)$ & $\delta(u)$ & $\frac{1}{\pi}K_0(|u|)$ & $\frac{1}{\sqrt{2}}\me^{-\sqrt{2}|u|}$ & $\frac{1}{\sqrt{2\pi}}\me^{-\frac{u^2}{2}}$ \\ 
        Asymptotic $u\rightarrow 0$ & - & $-\frac{1}{\pi}\ln{\frac{|u|}{2}}$ & - & $\frac{1}{\sqrt{2\pi}}$ \\ 
        Behaviour & Dirac delta & logarithmic singularity at $u=0$ & exponential & Gaussian
    \end{tabular}
    \caption{Prior distributions $P_\mathcal{U}(u)$ for particular choices of $\mu$. }
    \label{tab:values}
\end{table}

Asymptotically for large $u$, $P_\mathcal{U}(u)$ behaves as $P_\mathcal{U}\sim |u|^{\mu-1} \me^{-\sqrt{2\mu}|u|}$, showing that the tails are heavier than the Gaussian distribution. The asymptotic behaviour for $u\rightarrow 0 $, however, strongly depends on $\mu$; it is listed in Tbl.~\ref{tab:values} for specific values of $\mu$.

\bibliography{biblio}
\bibliographystyle{iopart-num}